\documentclass{article}

\usepackage[english]{babel}

\usepackage[letterpaper,top=2cm,bottom=2cm,left=3cm,right=3cm,marginparwidth=1.75cm]{geometry}

\usepackage{amsmath}
\usepackage{graphicx}
\usepackage[colorlinks=true, allcolors=blue]{hyperref}
\usepackage{authblk}

\title{Development of a Raman Lidar for the Southern Site of the Cherenkov Telescope Array Observatory}
\author{Patrick Brun}
\author{Omar Gabella}
\author{Stephane Rivoire}
\author{George Vasileiadis}
\affil{Laboratoire Univers et Particules de Montpellier, CNRS, Université de Montpellier, Montpellier, F-34095, France}
\begin{document}
\maketitle

\begin{abstract}
The future CTAO will reach a sensitivity and energy resolution never obtained
until now by any other high-energy gamma-ray experiment. Studying the atmospheric conditions
during the CTAO observations, namely the extinction and backscatter coefficients, permits a very
precise evaluation of the atmospheric UV absorption, a parameter that directly affects the energy
and flux spectra of the CTAO-studied sources. This paper describes the motivation for the LUPM
Raman Lidar system, a Lidar specifically designed to fulfill the CTAO requirements for the south
site in Chile. Preliminary results obtained during a two-year campaign at the OHP Observatory
are presented. Our results include estimations of the extinction, back-scatter, and Lidar ratio at
both 355 and 532 nm and confirm the conformity of our prototype to the CTAO requirements.
\end{abstract}

\section{Introduction}
The future Cherenkov Telescope Array Observatory (CTAO) [1, 2] will reach a sensitivity and energy
resolution never obtained until now by any other high-energy gamma-ray experiment. It is well-known
that atmospheric conditions contribute particularly in this aspect. Current Imaging Atmospheric
Cherenkov Telescope (IACT) experiments have pioneered several ways to introduce atmospheric calibration devices, 
among them the use of Lidars. Their experience has led to a coherent atmospheric
calibration strategy for the future CTAO observatory project. One of its key components consists of
the assessment of the atmospheric extinction profile throughout the entire path of Cherenkov photons
at two wavelengths that cover the sensitive regime of the light sensors used in the CTAO telescope
cameras. To achieve this, a Raman Lidar, operating at four distinctive wavelengths, namely 355/387
nm and 532/607 nm, will provide extinction profiles to continuously assess atmospheric extinction
across the observed science targets. This approach is sufficiently precise when either a single cloud
layer or no cloud layer is present in the observed field of view, a situation that is fulfilled practically
all the time when CTAO will be operative [3].
To precisely characterize extinction profiles up to 25-30 km height, a Raman Lidar should use
a powerful laser and a large reception mirror. These choices guarantee the highest precision at the
expense of strong interference with the CTAO observing telescope cameras, making it impossible
to simultaneously operate both. Since the telescopes need to repoint frequently, the Raman Lidar
must therefore be able to measure the extinction profiles within short time scales, typically a few
minutes. Considering the size of the optics, back-scattered light from very low altitudes, corresponding
to a few tenths of meters, became a big nuisance for the operation of the Lidar, saturating both the
photomultiplier and data acquisition operation chain. To overcome this issue, we have developed a
high-frequency gated photomultiplier electronic base that permits us to mask the operation of the
photomultiplier from those signals.
We will report here on the design and performance of the prototype Raman Lidar conceived at the
LUPM laboratory (Laboratoire Univers et Particules de Montpellier). Elastic and Raman spectra are
presented with a preliminary analysis of the obtained profiles. The use of a gated photomultiplier high
voltage base system demonstrates the benefits of such a development. We will refer to our system as
LRL (LUPM Raman Lidar). A similar project sharing the same scientific goals, based on the same mechanical structure but proposing different hardware solutions, can be found in \cite{BRL1,BRL2}. It is an ongoing project between several institutes and will be installed at the CTAO North site.
\section{The Instrument}
The Raman Lidar is optically and mechanically based on a telescope and mechanical assembly from the CLUE experiment \cite{CLUE}.  The building blocks of the Raman Lidar are identified as the container and telescope structure,  the optical emission and reception line, the polychromator, and the acquisition chain. We will detail each of them in the following sections. We will conclude by presenting some data acquired during observations done at the OHP Observatoire, where the Lidar is actually installed.
 
\subsection{Container and Telescope structure}

The  Lidar system is housed  in a container with outside dimensions of 1.7 m x 4.2 m x 2.1 m that can be opened like a clamshell during operation. Equipped with single-phase high-torque motors, both sidewalls can be opened/closed within 30 sec.,  {Figure 1}. The temperature inside the container can reach elevated values, especially in desert-type climates.  
\begin{figure}
	\centering
		\includegraphics[scale=0.47]{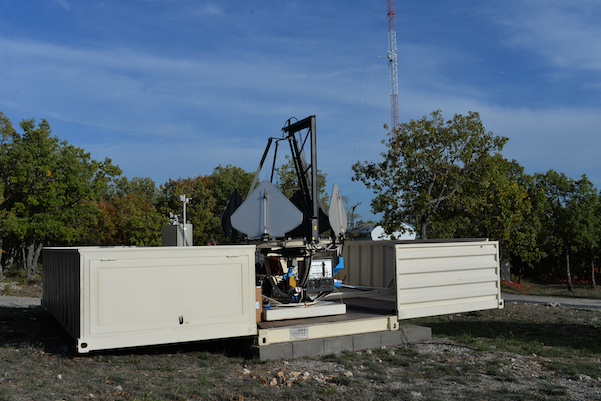}\includegraphics[scale=0.58]{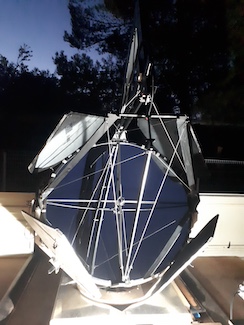}
	\caption{On the left are the CLUE experiment container and telescope structure used to build the LRL Lidar.  The photo shows the container in the operation position and the telescope pointing to Zenith.  On the right is the 1.8m telescope dish with the protective petals opened.}
	\label{FIG:1}
\end{figure}
A dedicated ventilation system ensures that air circulation inside the container keeps the temperature within operational limits of the equipment in place. The system requires a single phase 220 alternative  line  with a maximum current of 10A capacity  during normal operation.The total weight, including the Lidar system and the container, is approximately 3 tons.

Due to the small cross section of the Raman lines, about 2 to 3 orders of magnitude smaller  than the elastic cross section, a large reflecting area is needed to collect a sufficient amount of light within a restricted time lapse, normally 100 sec for the purposes of CTAO. The telescope,  {Figure 1}, is equipped  with a 1.8 m F/1 aluminized parabolic mirror. Measurements of the point spread function and reflectivity of the mirror showed that the mirror has maintained good optical quality, although the reflectivity has decreased to 64\%  at 350 nm, mostly due to aging, the original coating dated from 1985. Consequently, the mirror was  re-aluminized to reproduce its optimal performance (90\% reflectivity at all UV wavelengths).

  The telescope is mounted in an alt-azimuth configuration and can be pointed in any direction  employing  a pair of high speed stepper motors with a precision of 0.02 degrees. The maximum zenith angles available for pointing are dictated by the laser operation hazard rules, so they are limited to 65 degrees in our case. Altitude pointing is achieved using a belt driver system, while  the azimuth one is a direct coupling gear system. A motorized four petal Styrofoam pad element protects the mirror during non operating periods. 
The transmitter is  a military-spec Q-switched Nd:YAG laser from Quantel, referenced CFR400, with an output wavelength of 1064 nm. Harmonic modules at 532 and 355 nm provide the necessary frequencies for the elastic emission lines. The output pulse width is 5 ns, and the output energy is 330 mJ at 1064 nm. It is functioning at a repetition rate of 20 Hz with a beam divergence of  less than 1 mrad.  The laser emitting module is  bolted  on top of the telescope structure, and a set of 1 inch guiding mirrors exists for the laser beam just in front of the focal point of the telescope in a coaxial configuration.  
\begin{figure}
	\centering
		\includegraphics[scale=0.68]{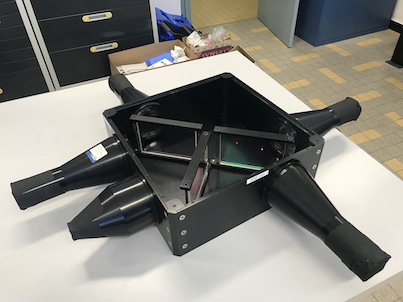},\includegraphics[scale=0.4465]{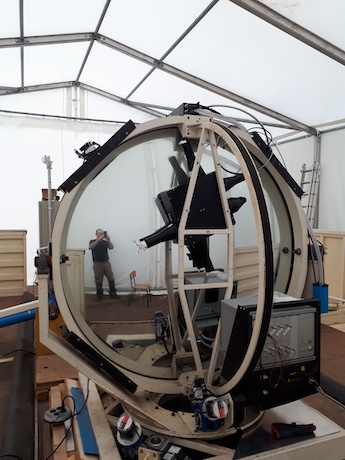}
	\caption{The polychromator unit (left-hand side photo) accommodating four read-out channels and relevant filters for separating the Elastic from the Raman lines. It is mounted on the back of the telescope structure and follows the movement of the telescope during pointing (right hand-side photo).}
	\label{FIG:2}
\end{figure}

Scattered  laser light from clouds and aerosols is gathered by the 1.8m mirror to  the focal point of the telescope at the entrance window of  a liquid fiber. We use a Lumatec 10mm in diameter fiber with a numerical aperture of 0.59 to guide the collected light towards the polychromator unit.   We use this kind of fiber to avoid stress effects due to bending during the pointing phase of the telescope while keeping an excellent UV transmission,  better than 70\%. 
\subsection{The Polychromator}
The output of the liquid fiber is fed to the input of the polychromator unit {Figure 2(L)}.  It is a custom designed and built unit by Raymetrics \cite{Ray} tailored to our needs. It spectrally separates the Raman components 387 nm and 607 nm and the Rayleigh-Mie components 355 nm and 532 nm via a set of notch splitters.  Each wavelength line is then focused on a dedicated  photomultiplier through a set of two custom designed Plano Convex lenses mounted in front of the photocathode that spreads the light signal uniformly.  A set of interference filters are mounted in-between to assure a measured leakage factor better than $10^{-9}$. We use an R329P 12 dynode photomultiplier from Hamamatsu  for all four lines.
\begin{figure}
	\centering
		\includegraphics[width=15.5cm,height=4cm]{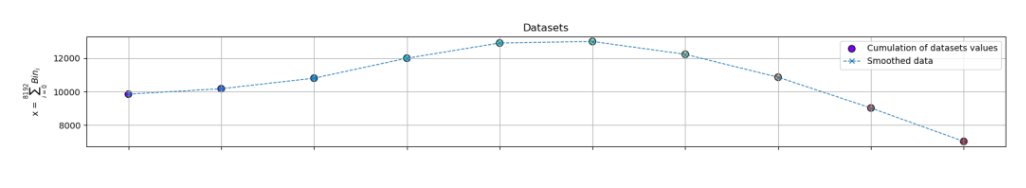},
        \includegraphics[width=15.5cm,height=4cm]{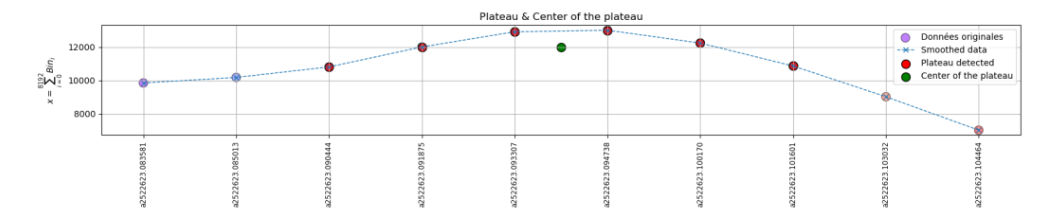}
	\caption{Output of the automatic alignment algorithm. The top and bottom plots correspond to the X and Y axis scans of the received signal. The whole procedure takes 1-2 min to conclude. The middle of the plateau detected corresponds to the most probable optimized position per axis. The  green dot corresponds to the selected position.}
	\label{FIG:2}
\end{figure}

The polychromator is designed to be lightweight, mechanically modular, and optically efficient. It is mounted on the back of the telescope dish, an integral part of the telescope structure. We opted for this solution to minimize any misalignment issues during the pointing phase of the Lidar, the stress on the liquid fiber, and also keep all signal tracks and cabling to a minimum. {Figure 2(R)}. 
\begin{figure}
	\centering
		\includegraphics[width=7.5cm,height=6cm]{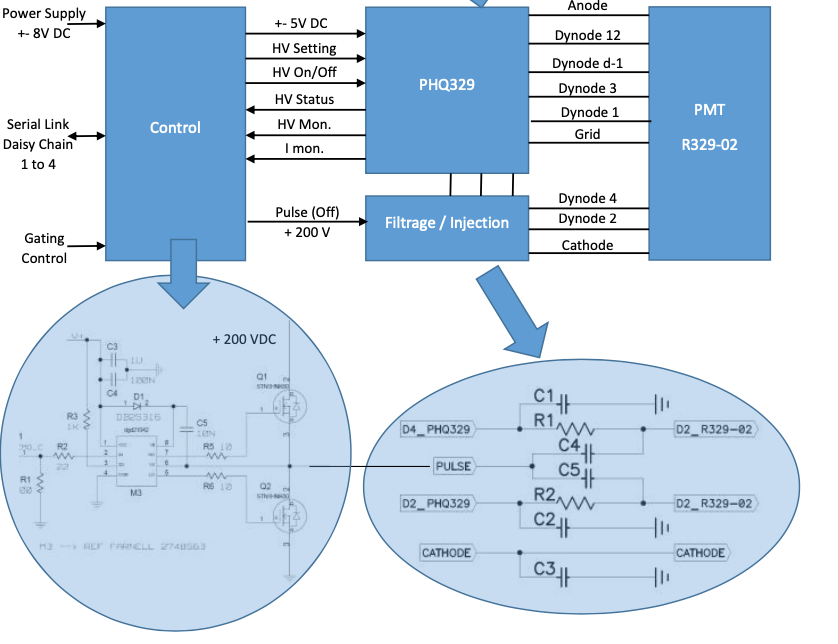},\includegraphics[width=7.5cm,height=6cm]{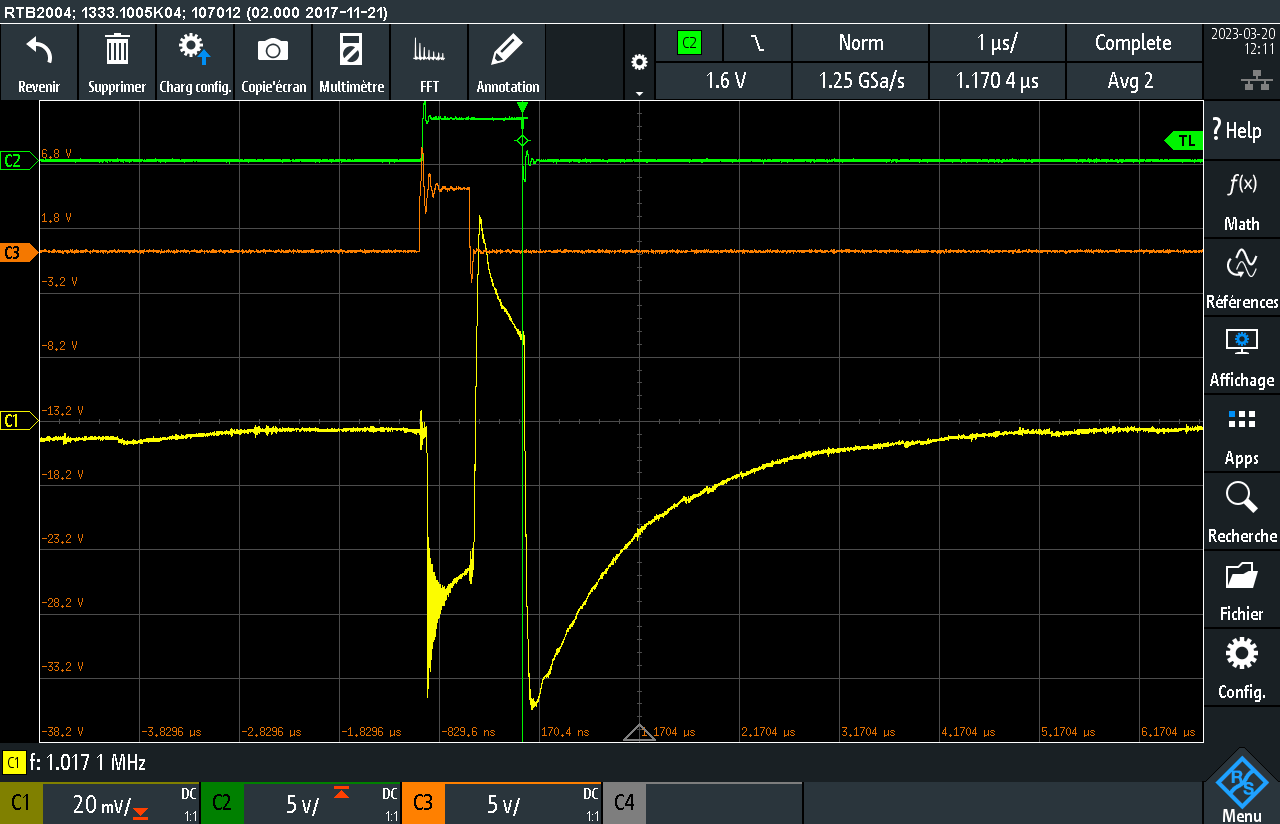}
	\caption{The photomultiplier gated schema was designed and built at LUPM. The right plot shows the concept of the trigger/gating logic and the way the high voltages are distributed on the photomultiplier base. The right-hand plot is the photomultiplier signal measured after a typical gate operation.}
	\label{FIG:3}
\end{figure}
\subsection{Automatic Alignment Protocol}
We have developed an automatic alignment system based on a CPU driven micro-motor assembly and motorized mirror so that the LIDAR reception can be optimized automatically with no need of human intervention.  The auto-alignment system procedure applied in this work is based on a method reported in \cite{pal}. A dedicated user driven application can align the system within 60 seconds to the required precision. The algorithm works in the principle  of maximizing the  received signal, within a certain range,  by tilting in a snail-wise movement the motorized alignment mirror. In Figure 3, some typical responses of the automatic procedure are shown. A plateau is expected on both axes when the alignment conditions are optimal. The green dot in the lower plot indicated the position selected in this particular case.

\subsection{The Acquisition chain}
A known issue operating Lidars in general is driving the acquisition chain to overcharged mode due to an excessive  amount of received light. This light is due either to the laser beam itself or, more importantly, to scattered laser pulses from  very low altitudes, a few tenths of meters. As a consequence, it drives the photomultiplier photocathodes to overload mode. This introduces slow recovery times, short-term instabilities, and after-pulsing. Thus, some photomultiplier overload protection is mandatory. Mechanical shutters are effective in preventing overload, but their response characteristics are too slow for short-range Lidars systems.  Optical choppers perform better but are still not adequate. Electro-optical modulators can be used as shutters, but they can be leaky and result in significant signal loss.  Electrical pulsing of the photomultiplier voltages seems to be the best available means of protecting against overload and was selected for this purpose in our design.

The pulsing technique adopted in our case, see Figure 4(L), is as follows:  A fast pulse lowers the first two photomultiplier high voltage base dynodes a few tenths of volts for an adjustable time interval until the overload  time is over.  The photomultiplier is thus in an OFF state.  The width of this pulse, corresponding to the OFF state, is user adjustable.  In our case, Zeemax simulations, performed by Raymetrics, demonstrated that due to optical aberrations signals received from altitudes below 600 meters would not fulfill the performance requirements. So the pulse width was set at this value by default.  After this time interval, the photomultiplier regains its normal operational status within 700 nsec.  

The waveforms of Figure 4(R) show the sampling pulse and the photomultiplier tube output.  We used normal observational conditions to obtain these data.  The gain turnoff  pulse was adjusted to an equivalent of 600 m for these tests, a value that permits us to observe at very low altitude with no interference from stray light coming from the laser.  No after-pulsing was observed during these preliminary tests, but still more field testing is necessary to evaluate the full potential of this technique. The  photomultiplier output signals are then fed to a standard industrial LICEL acquisition crate.  It is equipped with four TR20-12 bits modules, while additional boards ensure the synchronization of the laser trigger and DAQ. 

Data thus acquired are stored and analyzed with a software package we have developed specifically for this phase of the project. Eventually the CTAO-ACADA software protocol \cite{ACADA} will be implemented for the data transfer during the integration phase of the project.

\section{Data analysis and performance studies}
The advantage of the Raman Lidar technique is that it can measure extinction and back-scattering directly, eliminating the need for an assumption of the extinction-to-back-scatter ratios when solving the Lidar equation.  However, the only disadvantage of using the Raman Lidar technique is the relatively weak Raman scattering compared to the Rayleigh or fluorescence scattering, but this limitation can be overcome by the  use of high power lasers and/or long integration times. Thus, aerosol optical property profiles can be resolved in both time and space, enabling a vertical snapshot for the precise identification of aerosol distribution.

The LRL has been installed and operated at the Observatoire de Haute Provence (OHP) starting at the end of 2023 and until the end of 2025. This is an astronomical  site, where a number of optical telescopes are also installed and operational. The observing conditions were pretty optimal to evaluate the performance of our Lidar system.  The only drawbacks of the site were the extremely high levels of humidity, higher than 70\%, and very low temperatures, close or below zero during winter time, for more than half of the year, thus limiting the observing periods possible. Also the Zenith angle observations were exclusively done at 90 degrees due to air flight regulations. Several data acquisition campaigns were contacted, and we will report on some preliminary analysis of our data. The aim of these campaigns was mainly focused on determining operational issues with our design and obtain atmospheric profiles as close as possible to the CTAO requirements. Thus the acquisition time per profile was fixed to 100 sec and up to the  maximum altitude possible.

\subsection{Combined Elastic and Raman Method}
 Determining the vertical profile of the particle extinction and back-scatter coefficient at 355 nm and 532 nm relies on the Lidar inversion technique, and this is implemented following the reception of the nitrogen vibration Raman signals at 387 nm and 607 nm, as proposed by Ansmann et al \cite{ANSM}. The Lidar equation for the elastic scattering  signal is expressed as:
 \begin{equation}
P(\lambda_{L},z)=K_{\lambda L} \frac{O(z)}{z^2}  [\beta_{aer}(\lambda_{L},z)+\beta_{mol)}(\lambda_{L},z)]
\times  \textit{exp}  \{-\int^{z}_{0}[a_{aer}(\lambda_{L},z')+a_{mol}(\lambda_{L},z')]dz'\}
\end{equation}
and for the nitrogen Raman scatter, 
 \begin{equation}
 \begin{split}
P(\lambda_{R},z)=K_{\lambda R}  \frac{O(z)}{z^2}N_{R}(z)\frac{d\sigma_{\lambda_{R}}(\pi)}{d\Omega} 
\times  \textit{exp}  \{-\int^{z}_{0}[a_{aer}(\lambda_{L},z')\\+a_{mol}(\lambda_{L},z')+a_{aer}(\lambda_{R},z')+a_{mol}(\lambda_{R},z')]dz'\}
\end{split}
\end{equation}
where $P(\lambda_{R},z)$ are the return signals from distance z at the Raman wavelength of $\lambda_{R}$ O(z)  is the overlap correction factor,  $N_{R}(z)$ is the molecule number density of the Raman-active gas, and $d\sigma_{\lambda_{R}}/d\Omega$ is the range  independent differential Raman cross section for the backward direction. 
\begin{figure}
	\centering
		\includegraphics[scale=0.36]{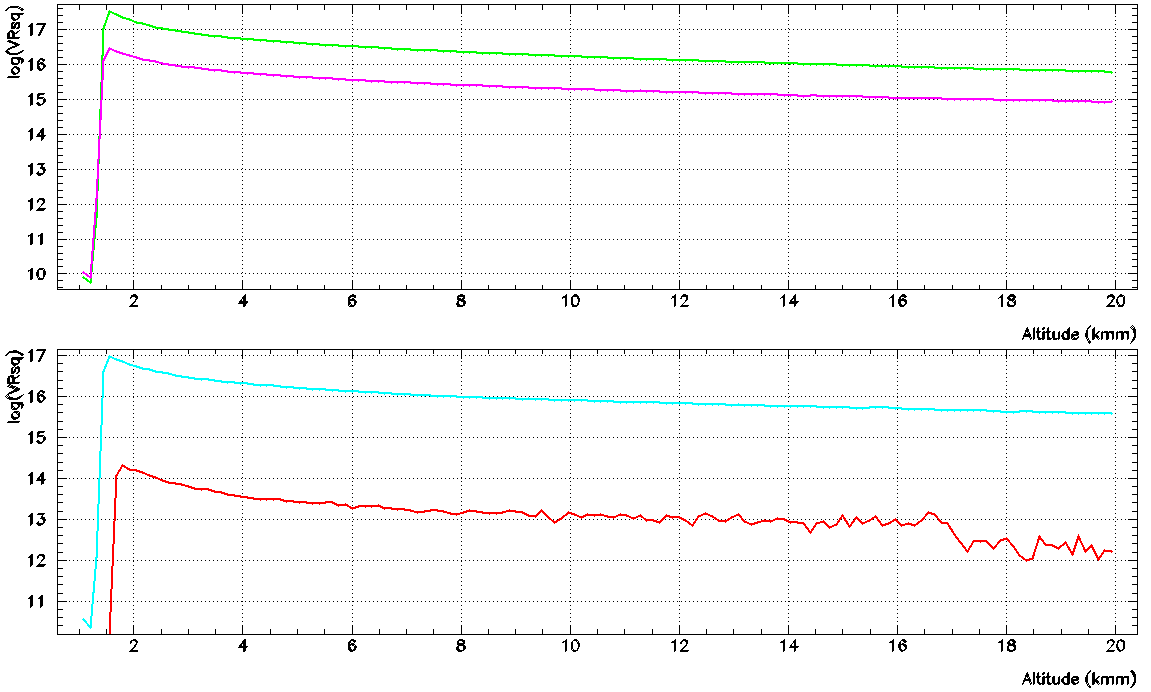}
	\caption{Range corrected return signals acquired during various data taking periods at OHP Observatory. The top plot corresponds to the 355 nm (green)/387 nm (magenta) channel, and the bottom one to the 532 nm (blue)/607 nm (red).}
	\label{FIG:4}
\end{figure}
The particle extinction coefficient can be determined as follows:
\begin{equation}
a_{aer}(\lambda_{L},z)=\frac{\frac{d}{dz}\left[ ln\frac{N_{R}(z)}{P(\lambda_{R},z)z^2}  \right] -a_{mol}(\lambda_{L},z)-a_{mol}(\lambda_{R},z)}{1+(\frac{\lambda_{L}}{\lambda_{R}})^k}
\end{equation}
In equation (3), Ångström’s law has been used while  \textit{k} is assumed to be one \cite{Ang}. It is foreseen though that k will be measured directly by CTAO using Moon and Sun photometers.

The first task, solving the calculation of the  extinction coefficient from aerosols, we continue with the calculation of the back-scatter coefficient. First, the elastic Lidar equation is normalized with its value at a specific reference height${z_0}$. Then the Raman Lidar equation is normalized with its value at the same reference height. Finally, the normalized elastic and Raman equations are divided. The described approach can be found described in more detail in the literature \cite{Ang,Ang2}. The advantage with this procedure is that it makes the power of the laser, the height independent parameters, and the overlap function cancel out. It is assumed that the overlap function is the same for the Raman and elastic Lidar equations. The aerosol back-scatter coefficient is then given by the following equation:
\begin{equation}
\beta_{aer}(\lambda_{L},z)=-\beta_{mol}(\lambda_{L},z)+\beta_{mol}(\lambda_{L},z)\frac{\frac{d}{dz}\left[ ln\frac{N_{R}(z)}{P(\lambda_{R},z)z^2}  \right] -a_{aer}(\lambda_{L},z)-a_{mol}(\lambda_{R},z)}{1+(\frac{\lambda_{L}}{\lambda_{R}})^k}
\end{equation}
To evaluate this equation, the reference height, $z_0$, needs to be set. The reference height is normally set such that the backscatter coefficient from
aerosols is much smaller than the back-scatter coefficient from molecules. Then the following approximation can be used, $(\beta_{aer}(\lambda_{0},z_{0})+\beta_{mol}(\lambda_{0},z_{0}))\approx \beta_{mol}(\lambda_{0},z_{0})$ and the back-scatter coefficient from aerosols at the reference height does not need to be known. 

Finally, the aerosol Lidar ratio can be calculated :
 \begin{equation}
LR_{aer}(\lambda_{L},z)=\frac{a_{aer}(\lambda_{L},z)}{\beta_{aer}(\lambda_{L},z)}
\end{equation}

\subsection{Data processing}
We start by producing the range corrected signals for all elastic and Raman channels.  The pre-processing of the raw Lidar data starts with spatial averaging. Spatial averaging is done to increase the signal-to-noise ratio of the Lidar measurements. For the present Lidar data analysis, we have aimed at a 30 m space-averaged data to increase the signal-to-noise ratio (SNR).  

We continue by calculating the background noise to be removed from all smoothed analog and photon counting (PC) signals. The background noise is inherent in the received signal, and it can be considered to be almost constant throughout the whole range in the signal. To remove the noise from the signals, first we have to calculate the noise offset value. For that, we select a region in the higher heights where aerosol content can be considered to be negligible and the back-scattered signal from such heights will mainly be due to the background radiation. The average of the Lidar signal in this altitude range can be considered as noise, and this is subtracted from the Lidar signal throughout the whole range. For our study, we have selected the area above 20 km and calculated the noise value for both the analog and PC channels and subtracted it from the respective signals.

Since our system is coaxial by design, calculating the distance to the full overlap range R \cite{ov1}, and following  similar considerations as on \cite{rlr1}, leads to a value of 50m or 95m without and with aberrations .  We  finalize  the signal processing by combining the analog and PC signals. The Lidar data collected in analog mode exhibits high linearity at lower heights, whereas the PC signal provides good sensitivity at higher altitudes. The SNR is high for analog at lower heights and for PC at higher heights. Due to this reason, there is a need to find a signal that shows higher SNR at both lower and higher heights. As a solution, we combine both the signals into a single signal through a process known as gluing. Although there are numerous strategies to achieve this, we have followed a method proposed by Licel itself \cite{Lic}.

Some typical nighttime profiles (range corrected) obtained after the processing phase described above are shown in Figure 5.  The top plot corresponds to the 355/387 nm elastic lines, while the bottom ones correspond to the 532/607 nm.  We can clearly distinguish the benefit of the photomultiplier gating protocol described beforehand. The profiles are free of saturation for altitudes as low as 600 m, a value predefined in our data acquisition for the specific observation conditions. Further optimizations will be needed, though, to achieve the maximum capabilities of this specific technique. On the other end of the spectrum, we were able to acquire reasonable profiles up to 25 km in altitude, but data corresponding to the 607 nm line were much less efficient. This is expected due to the reduced cross-section of this Raman line, so more optimizations will be needed to achieve the expected design goals. So we have concentrated our analysis efforts on altitudes up to 10 km, where the response of the 607 nm was judged acceptable.
\begin{figure}
	\centering
	\includegraphics[width=13.5cm,height=11cm]{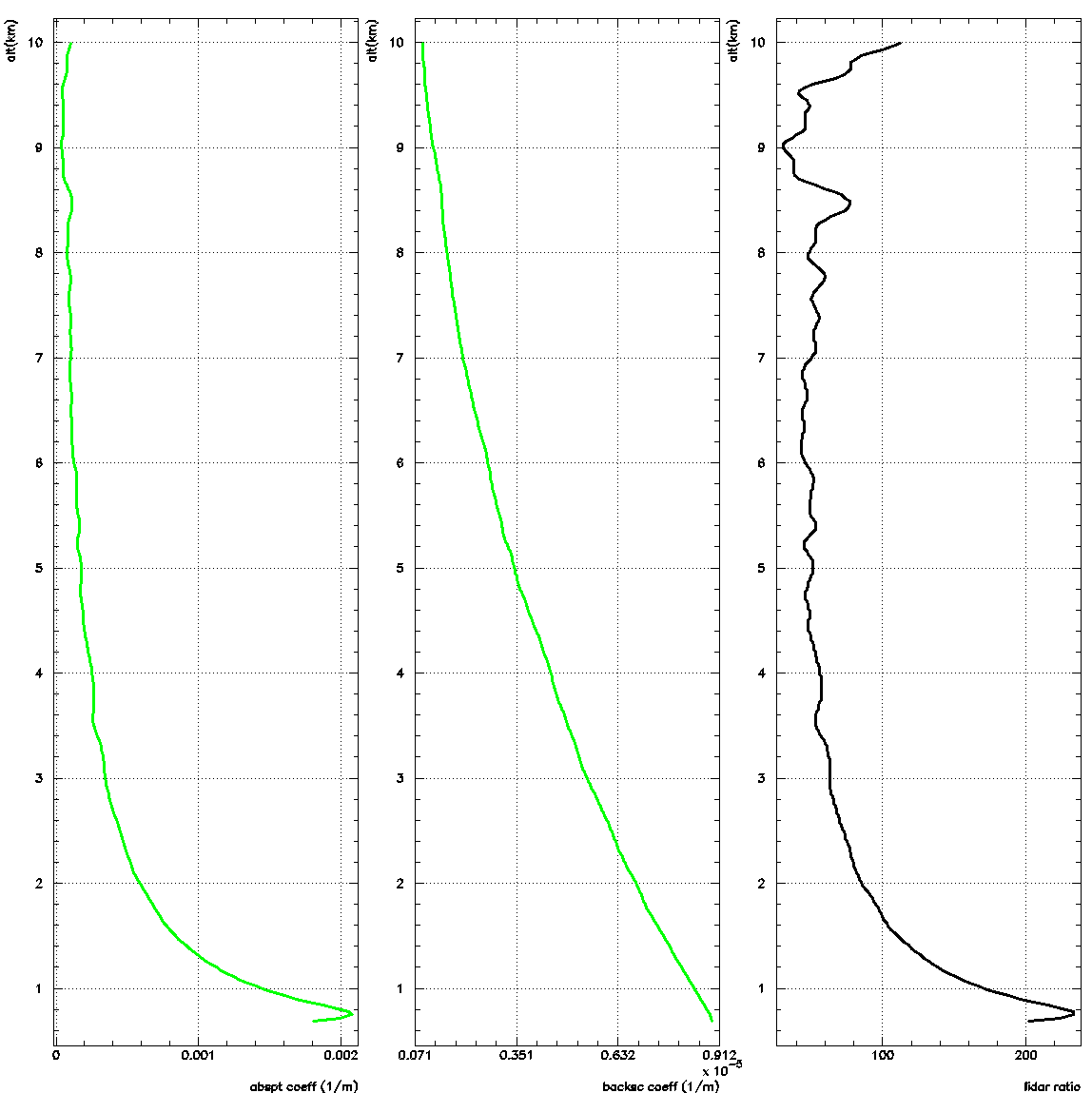}
    \includegraphics[width=13.5cm,height=11cm]{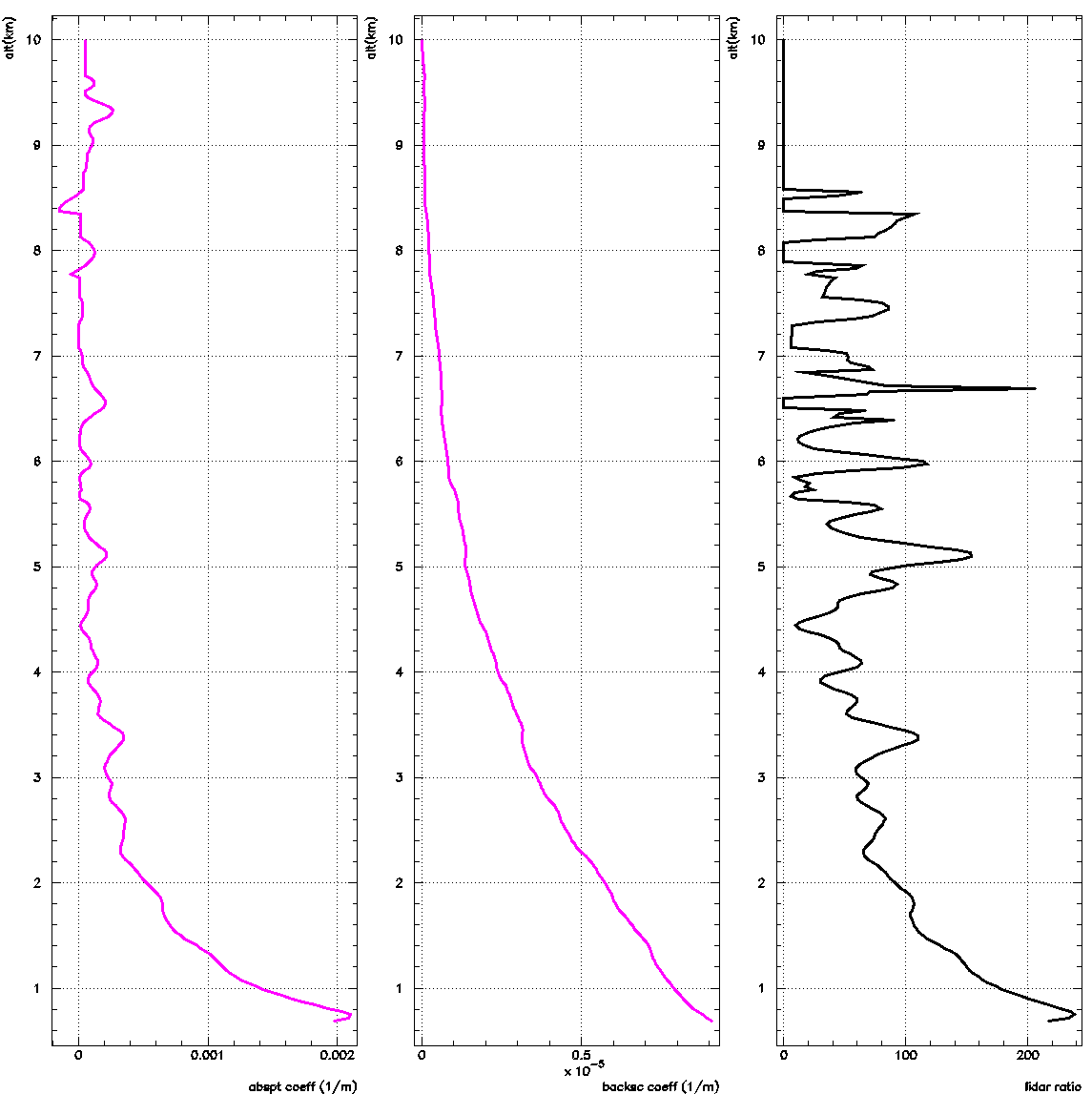}
	\caption{Calculation of the extinction and back-scatter coefficients for the 355 nm (in green)  and 532 nm (in magenta) lines.  The corresponding lidar ratios are also shown.
}
	\label{FIG:5}
\end{figure}
\subsection{Results and discussions}
We continue by calculating the aerosol extinction and back-scatter coefficients for both the 355 nm and 532 nm lines based on the formulation presented above (3 and 4). One of the big problems when trying to solve the Raman Lidar equation with the measured data is that it has a differentiation of the measured Raman signal. Since the Raman signal is a photon counted signal, the amount of filtering in the previous step is important. Numerical differentiation can be done in many ways. The most obvious is to take the derivative between two samples at a time. The problem with this is that the derivative will be too noisy. To suppress the noise factor, another solution is chosen here. The differentiation is made with a linear least squares fit to several measurement points. The slope in the fitted data is taken as the derivative.  After this, the last part of equation (3) is applied:  subtraction of the extinction coefficients from molecules and division with the wavelength dependence. The aerosol back-scatter coefficient is calculated afterward  using equation (4).  A reference height has been chosen at a height that corresponds to the absence of aerosols. For this analysis it has been chosen at 15 km.

Typical profiles  obtained are shown in Figure 6. Results shown were acquired on the 24th of February at 23h05 with temperatures close to zero and humidity at the 30-40\% level. Skies were clear. The results obtained for the 355/387 nm lines are more consistent due to better statistics of the Raman line. The acquired spectrum for the 607 nm line was the one that presented the most difficulties to obtain; thus, the corresponding extraction of the extinction coefficient is much more noisy for higher altitudes. Also, as  mentioned before, the gluing procedure was not optimized at this stage, and that explains partially the results obtained at higher altitudes. Lidar radio values are also shown, in the range of 25-100, that are consistent with what we've expected for the specific site and weather conditions, Mediterranean south winter conditions.   
\section{Conclusions}
The LRL prototype, developed specifically for the needs of the CTAO observatory, is reported here. A four channel spectrometer permits the calculation of the extinction profiles up to 10 km simultaneously for the 355 nm and 532 nm lines. The introduction of a triggered readout system, developed in our laboratory, permits us to obtain results for altitudes as low as a few hundred meters, 600  in our case. After a successful 20-month test campaign at the OHP site, we were able to extract the extinction parameters and Lidar ratio for both datasets and in relatively close accordance with archived EARLINET data for similar conditions. Near future development will focus on better implementing the gluing algorithms for better high altitude performance. Upon acceptance of the project by the CTAO management, the LRL Lidar will be installed permanently at the south site of CTAO, in Chile, on a date expected to be around the end of 2027. 
\section{Acknowledgements}
This work was conducted in the context of the CTAO Consortium.
{}
\end{document}